\documentclass[journal]{IEEEtran}
\usepackage{amsmath,amsfonts}
\usepackage{algorithmic}
\usepackage{array}
\usepackage{subfig}
\usepackage{textcomp}
\usepackage{stfloats}
\usepackage{url}
\usepackage{verbatim}
\usepackage{graphicx}
\def\BibTeX{{\rm B\kern-.05em{\sc i\kern-.025em b}\kern-.08em
    T\kern-.1667em\lower.7ex\hbox{E}\kern-.125emX}}
\usepackage{balance}

\usepackage{amsmath}
\usepackage{amssymb}
\usepackage{color}
\usepackage{xspace}

\usepackage{multirow}
\usepackage{tabularx}
\usepackage{tablefootnote}
\usepackage{booktabs}
\usepackage{makecell}
\usepackage{bbding}

\usepackage{amssymb}
\usepackage{pifont}

\usepackage{microtype}
\usepackage{graphicx}
\usepackage[table]{xcolor}    
\usepackage{booktabs}
\usepackage{hyperref}
\usepackage{enumitem}

\begin{document}
\title{The CCF AATC 2025 Speech Restoration\\ Challenge: A Retrospective}
\author{
Junan Zhang,~\IEEEmembership{}
Mengyao Zhu,~\IEEEmembership{}
Xin Xu,~\IEEEmembership{}
Hui Bu,~\IEEEmembership{}
Zhenhua Ling,~\IEEEmembership{}
Zhizheng Wu~\IEEEmembership{}

\thanks{J. Zhang and Z. Wu are affiliated with The Chinese University of Hong Kong, Shenzhen. e-mail: \href{mailto:junanzhang@link.cuhk.edu.cn}{junanzhang@link.cuhk.edu.cn}}
\thanks{M. Zhu is affiliated with Audio Department, Huawei CBG.}
\thanks{X. Xu and H. Bu are affiliated with Beijing AISHELL Technology Co., Ltd.}
\thanks{Z. Ling is affiliated with University of Science and Technology of China.}
\thanks{Z. Wu is the corresponding author. e-mail: \href{mailto:wuzhizheng@cuhk.edu.cn}{wuzhizheng@cuhk.edu.cn}}
}

\maketitle

\begin{abstract}
Real-world speech communication is rarely affected by a single type of degradation. Instead, it suffers from a complex interplay of acoustic interference, codec compression, and, increasingly, \textit{secondary artifacts} introduced by upstream enhancement algorithms. To bridge the gap between academic research and these realistic scenarios, we introduced the CCF AATC 2025 Speech Restoration Challenge. This challenge targets \textit{universal blind speech restoration}, requiring a single model to handle three distinct distortion categories: acoustic degradation, codec distortion, and secondary processing artifacts. In this paper, we provide a comprehensive retrospective of the challenge, detailing the dataset construction, task design, and a systematic analysis of the 25 participating systems. We report three key findings that define the current state of the field: (1) \textit{Efficiency vs. Scale:} Contrary to the trend of massive generative models, top-performing systems demonstrated that lightweight discriminative architectures ($<$10M parameters) can achieve state-of-the-art performance, balancing restoration quality with deployment constraints. (2) \textit{The Generative Trade-off:} While generative and hybrid models excel in theoretical perceptual metrics, breakdown analysis reveals they suffer from ``reconstruction bias'' in high-SNR codec tasks and struggle with hallucination in complex secondary artifact scenarios. (3) \textit{The Metric Gap:} Most critically, our rank correlation analysis exposes a strong negative correlation ($\rho=-0.8$) between widely-used reference-free metrics (e.g., DNSMOS) and human MOS when evaluating hybrid systems. This indicates that current metrics may over-reward artificial spectral smoothness at the expense of perceptual naturalness. This paper aims to serve as a reference for future research in robust speech restoration and calls for the development of next-generation evaluation metrics sensitive to generative artifacts.
\end{abstract}


\section{Introduction}
\label{sec:intro}

\begin{figure}[!htbp]
  \centering
  \includegraphics[width=0.95\linewidth]{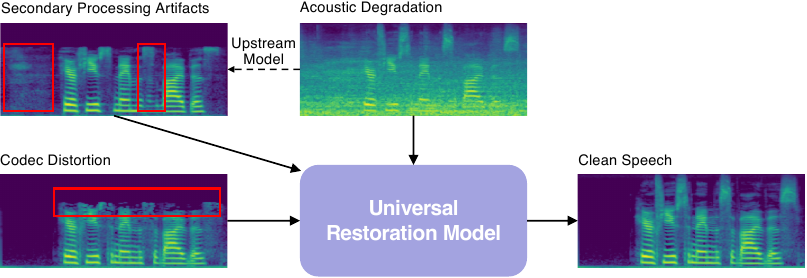}
  \caption{Overview of the universal blind speech restoration challenge. A single model must recover clean speech from three degradation types without knowing the distortion source: (1) Acoustic degradation from noise and reverberation. (2) Secondary processing artifacts created by passing degraded speech through an upstream enhancement model (dashed arrow), with red boxes indicating residual noise and spectral artifacts. (3) Codec distortion, where red boxes highlight high-frequency cutoff and quantization noise from mp3 compression.}
  \label{fig:overview}
\end{figure}

\begin{table*}[htbp]
  \small
  \centering
  \caption{Comparison of existing challenges and representative methods. Unlike previous benchmarks focusing on isolated degradations, CCF AATC 2025 targets universal restoration across acoustic, codec, and secondary artifacts.}
  \label{tab:related-work}
  \begin{tabular}{ccccc}
      \toprule
      \textbf{Type} & \textbf{Name} & \textbf{Acoustic Degradation} & \textbf{Codec Distortion} & \textbf{Secondary Processing Artifacts} \\
      \midrule
      \multirow{5}{*}{Challenge}
      & DNS Challenge~\cite{reddy2021icassp} & \ding{51} & \ding{55} & \ding{55} \\
      & URGENT Challenge 2024~\cite{zhang24h_interspeech} & \ding{51} & \ding{55} & \ding{55} \\
      & URGENT Challenge 2025~\cite{saijo2025urgent} & \ding{51} & \ding{51} & \ding{55} \\
      \cmidrule(lr){2-5}
      & CCF AATC 2025: Speech Restoration & \ding{51} & \ding{51} & \ding{51} \\
      \midrule
      \multirow{6}{*}{Method}
      & Voicefixer~\cite{liu2022voicefixer} & \ding{51} & \ding{55} & \ding{55} \\
      & AnyEnhance~\cite{zhang2025anyenhance} & \ding{51} & \ding{55} & \ding{55} \\
      & Apollo~\cite{li2025apollo} & \ding{55} & \ding{51} & \ding{55} \\
      & Diffiner~\cite{sawata2023diffiner} & \ding{55} & \ding{55} & \ding{51} \\
      & SpeechRefiner~\cite{li2025speechrefiner} & \ding{55} & \ding{55} & \ding{51} \\
      \bottomrule
  \end{tabular}
\end{table*}

Speech restoration, the process of recovering clean speech from degraded signals, is a cornerstone of modern audio engineering. While significant progress has been made in isolated sub-tasks such as denoising~\cite{reddy2021icassp}, dereverberation~\cite{richter2023sgmse}, and bandwidth extension~\cite{li2025apollo}, real-world communication pipelines present a far more complex landscape. Speech signals in the wild typically suffer from a cascade of distortions~\cite{liu2022voicefixer,zhang24h_interspeech,zhang2025anyenhance,wang2025metis,li2024masksr,zhang2025multi}, including acoustic interference, lossy compression, and upstream processing errors.

Current benchmarks, as summarized in Table~\ref{tab:related-work}, largely focus on subsets of these problems. The DNS Challenges~\cite{reddy2021icassp} primarily target acoustic degradation, while recent efforts like the URGENT Challenge 2025~\cite{saijo2025urgent} have begun to incorporate codec artifacts. However, a critical gap remains: the prevalence of \textbf{Secondary Processing Artifacts}. In modern Voice over Internet Protocol (VoIP) and conferencing systems, audio is often pre-processed by imperfect enhancement algorithms (e.g., noise gates or legacy denoisers) before transmission, introducing non-linear artifacts such as musical noise, spectral holes, and phase discontinuities. Restoring such processed speech requires a model to distinguish between natural speech features and artificial algorithmic residues—a capability that existing benchmarks have yet to systematically evaluate.

To bridge this gap, we introduced the \textbf{CCF AATC 2025 Speech Restoration Challenge}\footnote{\url{https://ccf-aatc.org.cn/}}. This challenge defines a novel task: \textbf{Universal Blind Speech Restoration}. As illustrated in Fig.~\ref{fig:overview}, participants must develop a single model capable of handling three distinct input streams—(1) Acoustic Degradation, (2) Codec Distortion, and (3) Secondary Processing Artifacts—without prior knowledge of the distortion type. This setup pushes the boundaries of model generalization, requiring systems to balance smoothing (for denoising) with texture generation (for codec recovery) and artifact removal.

In this paper, we provide a comprehensive retrospective of the challenge, analyzing the submissions from 25 participating teams. These systems represent a diverse cross-section of current methodologies, including Time-Frequency (T-F) domain discriminative models, masked generative models, and emerging hybrid architectures (e.g., gan-score model-based integration). Beyond a simple leaderboard, our systematic analysis yields three critical insights that define the current state of the field:

\begin{itemize}
    \item \textbf{Efficiency vs. Scale:} Contrary to the prevailing trend of massive generative models, our results demonstrate that lightweight discriminative architectures (parameters $<10$M) dominate the top rankings. These models offer a superior trade-off between restoration fidelity (Word Accuracy) and computational cost, proving their viability for edge deployment.
    \item \textbf{The Generative Trade-off:} While generative and hybrid models excel in generating perceptually smooth spectra, breakdown analysis reveals a ``reconstruction bias.'' In high-SNR tasks like Codec Restoration, generative models often aggressively resynthesize the signal, leading to lower phase fidelity (PESQ) and intelligibility compared to discriminative approaches.
    \item \textbf{The Metric Gap:} Most critically, we expose a severe misalignment in evaluation metrics. Our correlation analysis reveals a strong \textbf{negative correlation} ($\rho = -0.8$) between widely used reference-free metrics (e.g., DNSMOS) and human subjective ratings (MOS) when evaluating hybrid systems. This indicates that current metrics over-reward artificial spectral smoothness at the expense of perceptual naturalness, calling for a paradigm shift in how we evaluate generative speech enhancement.
\end{itemize}

This paper aims to serve not only as a summary of the CCF AATC 2025 challenge but also as a reference for future research in robust, universal speech restoration, highlighting the need for next-generation metrics sensitive to generative artifacts and improved generative model design.

\section{Challenge Overview}
\label{sec:task}

\subsection{Problem Formulation}
The goal of this challenge is \textbf{Universal Blind Speech Restoration}. Let $x \in \mathbb{R}^{T}$ denote a mono clean speech waveform of length $T$. In real-world transmission, $x$ can be corrupted by three distinct degradation mechanisms: acoustic degradation, codec distortion, and secondary processing artifacts. The observed degraded signal $y$ is modeled as:
\begin{equation}
y = \begin{cases} 
\mathcal{D}_{\text{ac}}(x, n, h) & \text{Type I: Acoustic} \\
\mathcal{D}_{\text{codec}}(x, r) & \text{Type II: Codec} \\
\mathcal{M}_{\theta}(\mathcal{D}_{\text{ac}}(x, n, h)) & \text{Type III: Secondary}
\end{cases}
\end{equation}
where $\mathcal{D}_{\text{ac}}$ represents the acoustic degradation, $n$ is the noise, $h$ is the room impulse response; $\mathcal{D}_{\text{codec}}$ represents codec distortion and $r$ is the bitrate; and $\mathcal{M}_{\theta}$ denotes a non-linear upstream enhancement model with parameters $\theta$. Note that Type III degradation is composite, applying algorithmic artifacts onto acoustically degraded speech. Specifically the three degradation types are:

\begin{itemize}
    \item \textbf{Acoustic Degradation:} We considered four types of acoustic degradation: reverberation, clipping, bandwidth limitation, and noise. To generate the acoustic degraded speech: given clean speech $x$, room impulse response $h$, and noise $n$, the acoustic degraded speech is generated as:
    $y = \mathcal{D}_{\text{ac}}(x, n, h) = \text{LPF}(\text{Clip}(x \ast h)) + n$, where $\text{LPF}$ is the low-pass filter, $\text{Clip}$ is the clipping function.
    \item \textbf{Codec Distortion:} We applied MP3 compression to the audio signals to emulate artifacts from transmission over band-limited channels. $y = \mathcal{D}_{\text{codec}}(x, r) = \mathcal{E}_{\text{mp3}}(x, r)$, where $r$ is the bitrate.
    \item \textbf{Secondary Processing Artifacts:} We processed the acoustically degraded speech with a suite of speech enhancement models and used their outputs as an additional source of distortion: $y = \mathcal{M}_{\theta}(\mathcal{D}_{\text{ac}}(x, n, h))$, where $\theta$ is the parameters of the enhancement model.
\end{itemize}

The challenge requires learning a single restoration function $f_{\phi}: \mathbb{R}^{T} \to \mathbb{R}^{T}$ such that $\hat{x} = f_{\phi}(y)$ approximates $x$ with high fidelity, without prior knowledge of the distortion type. More details about the dataset construction are provided in Section~\ref{sec:degradation_pipeline}.

\subsection{Degradation Pipeline}
\label{sec:degradation_pipeline}
We implemented a customizable pipeline\footnote{\url{https://github.com/viewfinder-annn/anyenhance-v1-ccf-aatc/tree/main/data_simulation}} to simulate the three distortion types defined in Eq. (1).

\subsubsection{Acoustic Degradation ($\mathcal{D}_{\text{ac}}$)} 
To simulate realistic recording environments, we introduced acoustic distortions using the degradation pipeline from AnyEnhance~\cite{zhang2025anyenhance}. The pipeline consists of four stages:
\begin{itemize}
    \item \textbf{Reverberation:} ($p=0.5$) Convolution with RIRs $h$.
    \item \textbf{Clipping:} ($p=0.25$) Mimics microphone saturation with quantile thresholds between $[0.06, 0.9]$.
    \item \textbf{Bandwidth Limitation:} ($p=0.5$) Applies low-pass filtering to simulate diverse capture devices. Cutoff frequencies are sampled from $\{4, 8, 16, 24, 32\}$ kHz using various resampling kernels (Kaiser-best, Kaiser-fast, Polyphase).
    \item \textbf{Additive Noise:} ($p=0.9$) Mixes background noise $n$ with SNR sampled uniformly from $[-5, 20]$ dB.
\end{itemize}

\subsubsection{Codec Distortion ($\mathcal{D}_{\text{codec}}$)}
To simulate transmission artifacts over band-limited channels, we applied \textbf{mp3 compression} using the torchaudio backend. We sampled bitrates from a wide range: $\{24, 32, 48, 64, 96, 128\}$ kbps. This introduces characteristic artifacts such as high-frequency cutoff and quantization noise, particularly at lower bitrates (24-32 kbps).

\subsubsection{Secondary Processing Artifacts ($\mathcal{D}_{\text{sec}}$)}
This track simulates the artifacts introduced by imperfect upstream enhancement. We selected 10 representative speech enhancement models, including discriminative and generative models. The discriminative models included \textbf{Demucs}~\cite{defossez2019demucs}, a popular real-time waveform-to-waveform model; \textbf{FRCRN}~\cite{zhao2022frcrn}, which boosts feature representation through frequency recurrence; \textbf{VoiceFixer}~\cite{liu2022voicefixer}, a two-stage general restoration model; \textbf{NSNet2}~\cite{reddy2021icassp}, the lightweight baseline from the ICASSP 2021 DNS Challenge; and \textbf{TF-GridNet}~\cite{wang2023tf}, the time-frequency domain baseline from the Interspeech 2024 Urgent Challenge. The generative models consisted of \textbf{SGMSE+}~\cite{richter2023sgmse}, a versatile score-based generative model; \textbf{StoRM}~\cite{lemercier2023storm}, a hybrid model combining a predictive stage and a generative stage; \textbf{AnyEnhance}~\cite{zhang2025anyenhance}, a unified masked generative model with prompt-guidance and self-critic; \textbf{MaskSR}~\cite{li2024masksr}, a masked generative model that predicts acoustic tokens; and \textbf{LLaSE-G1}~\cite{kang2025llase}, a LLaMA-based model designed to perform multiple enhancement tasks.

\subsection{Dataset Construction}
\label{sec:dataset}
To support the challenge, we curated a large-scale training dataset and provided a development set sampled at \textbf{44.1 kHz}. All data was synthesized from high-quality, clean speech corpora to create corresponding degraded versions through the degradation pipeline described in Section~\ref{sec:degradation_pipeline}.

\subsubsection{Clean Speech Corpora}
The source speech is derived from three high-quality corpora. We strictly excluded specific speakers for development sets:
\begin{itemize}[leftmargin=*]
    \item \textbf{VCTK}~\cite{yamagishi2019vctk}: A multi-speaker English dataset with various accents. Excluded speakers \textit{p229, p236, p310}.
    \item \textbf{AISHELL-3}~\cite{AISHELL3}: A large-scale Mandarin Chinese speech corpus. Excluded speakers \textit{SSB0197, SSB0686, SSB1684}.
    \item \textbf{EARS}~\cite{richter2024ears}: A high-quality English speech dataset recorded in controlled environments. Excluded speakers \textit{p031, p065, p066}.
\end{itemize}

\subsubsection{Noise and RIR Data}
To simulate diverse acoustic environments, we utilized noise recordings from MUSAN~\cite{snyder2015musan}, the Urgent Challenge subset (DNS5+Wham)~\cite{zhang24h_interspeech}, FSD50K~\cite{fonseca2021fsd50k}, and DESED~\cite{turpault2019sound}. Room Impulse Responses (RIRs) comprised 62,668 samples from SLR26, SLR28, and self-collected rirs from AnyEnhance~\cite{zhang2025anyenhance}.

The training set was statically generated to ensure reproducibility. It contains 153,475 clean utterances, totaling $\sim$300 hours. For every clean utterance $x_i$, we generated three corresponding degraded versions $\{y_{i}^{\text{ac}}, y_{i}^{\text{codec}}, y_{i}^{\text{sec}}\}$. The development set consists of 500 utterances from excluded speakers, with the same distortion configuration as the training set, with 200 Acoustic, 100 Codec, and 200 Secondary.

\subsection{Baseline System}
We provided an official baseline system based on the AnyEnhance framework~\cite{zhang2025anyenhance}. The baseline is a lightweight, non-causal version of a masked generative model. It consists of a semantic enhancement stage and an acoustic enhancement stage, designed to progressively predict the acoustic tokens of clean speech from the degraded speech. The baseline model consists of 10 transformer layers (512 hidden dimension), resulting in 45.68M parameters. It predicts acoustic tokens of DAC tokenizer~\cite{kumar2024high}, so requires DAC's decoder to decode the tokens into speech. So the baseline model consists of 45.68M parameters for the generative backbone and 36.50M parameters for the DAC decoder, resulting in a total inference size of \textbf{82.18M}. The complete source code, pre-trained model, and detailed instructions for training and inference were made publicly available in a GitHub repository\footnote{\url{https://github.com/viewfinder-annn/anyenhance-v1-ccf-aatc}}.

\begin{table*}[!ht]
  \centering
  \caption{Evaluation metrics and scoring criteria for the challenge.}
  \small
  \label{tab:evaluation_criteria}
  \begin{tabular}{p{2.5cm} p{4cm} p{6cm} c}
    \toprule
    \textbf{Primary Metric} & \textbf{Secondary Metric (Weight)} & \textbf{Requirements} & \textbf{Max Score} \\
    \midrule
    \multirow{9}{*}{Objective Score} 
      & Objective Metrics (40\%) & Composite score of WAcc, DNSMOS, and PESQ with weights: 2:1:1. \newline
      Score by rank: 40 points for first rank, linearly decreasing to 16 points for last rank. & 40 \\
    \cmidrule(lr){2-4}
      & Parameter Count (20\%) & Scoring by model parameters: \newline
        $<$10M: 20 points \newline
        10M--20M: 16 points \newline
        20M--50M: 12 points \newline
        50M--100M: 10 points \newline
        $>$100M: 8 points & 20 \\
    \midrule
    \multirow{7}{*}{\makecell[c]{Subjective Score \\ (Final Round only)}}
      & Solution Innovation (20\%) & Original progress in algorithm/implementation: 10--20 points. \newline
        Optimization of existing methods: 0--10 points. & 20 \\
    \cmidrule(lr){2-4}
      & Listening Test (20\%) & Submissions ranked by subjective listening tests (MOS), divided into 4 tiers: \newline
        Tier 1: 20 points, Tier 2: 16 points, Tier 3: 12 points, Tier 4: 8 points & 20 \\
    \bottomrule
  \end{tabular}
\end{table*}

\subsection{Benchmarking Methodology}
\label{sec:benchmarking}

To benchmark the competing systems, we designed a multi-stage evaluation protocol that balances restoration quality, intelligibility, and computational efficiency. The competition is divided into a Preliminary Round (objective-only) and a Final Round (subjective and innovation assessment). The detailed scoring criteria are summarized in Table~\ref{tab:evaluation_criteria}.

\subsubsection{Preliminary Round: Objective Benchmarking}
The preliminary round serves as a quantitative filter. Teams are ranked based on a composite Objective Score, derived from two key components:

\begin{itemize}
    \item \textbf{Performance Metrics (40 points):} We employ a composite ranking system based on three complementary metrics to assess different aspects of restoration quality:
    \begin{itemize}
        \item \textbf{WAcc (Word Accuracy):} Measures speech intelligibility. We prioritize this metric with a 50\% weight (Ratio 2:1:1 against others) to penalize generative models that produce high-quality but hallucinated content. We use Whisper-large-v3\footnote{\url{https://huggingface.co/openai/whisper-large-v3}} as the ASR model.
        \item \textbf{DNSMOS}~\cite{reddy2021icassp}: Estimates perceptual quality (SIG, BAK, OVRL) without a reference signal, accounting for 25\% weight.
        \item \textbf{PESQ}~\cite{pesq2}: Measures waveform fidelity and spectral magnitude consistent with the reference, accounting for 25\% weight.
    \end{itemize}
    Teams are ranked on each metric individually, and the weighted average rank determines the final metric score. The evaluation script is available in the baseline repository\footnote{\url{https://github.com/viewfinder-annn/anyenhance-v1-ccf-aatc/blob/main/evaluation}}.
    \item \textbf{Efficiency Constraints (20 points):} To bridge the gap between academic research and industrial deployment, we explicitly evaluate model complexity. We enforce a ``lightweight first'' policy, where models are categorized into tiers based on parameter count. As shown in Table~\ref{tab:evaluation_criteria}, models under 10M parameters receive the maximum score, while those exceeding 100M receive the minimum. This design discourages the brute-force scaling of generative backbones and encourages efficient architectural innovations.
\end{itemize}

\subsubsection{Final Round: Subjective and Innovation Assessment}
The top six teams from the preliminary round advance to the finals. The evaluation expands to include human perception and algorithmic novelty:

\textbf{1) Subjective Listening Test (20 points):}
Given the distinct perceptual characteristics of different distortions, our subjective evaluation focuses on the \textbf{Acoustic Degradation} subset, which exhibits the highest perceptual variance. We recruited listeners to rate the restored audio using the Mean Opinion Score (MOS) standard.

\textbf{2) Solution Innovation (20 points):}
A panel of experts evaluates the technical novelty of the proposed solutions based on the submitted technical reports and code. Points are awarded for original architectural improvements (e.g., novel attention mechanisms, hybrid integration strategies) rather than simple hyper-parameter tuning or ensembling of existing models.

The final ranking is determined by the sum of the Preliminary Objective Score and the Final Round scores (Subjective + Innovation).

\section{Results and Analysis}
\label{sec:results}

In this section, we present a systematic evaluation of the participating systems. We first detail the experimental setup, followed by a holistic analysis of performance versus efficiency. Finally, we conduct a breakdown analysis across distortion types and investigate the correlation between objective metrics and human perception.

\subsection{Experimental Setup}
\label{subsec:setup}

\textbf{Test Dataset and Evaluation Protocol}
To assess generalization, we constructed a blind test set of 300 utterances using held-out internal clean speech mixed with noise from the TUT Urban Acoustic Scenes 2018 dataset~\cite{mesaros2018multi} and unseen RIRs. The corresponding degradation types are divided into three categories: (1) \textbf{Acoustic Degradation} (150 files); (2) \textbf{Codec Distortion} (50 files) across various bitrates; and (3) \textbf{Secondary Artifacts} (100 files) derived from the ten upstream models described in Section~\ref{sec:dataset}. The testset is available at \footnote{\url{https://github.com/viewfinder-annn/AnyEnhance-v1/releases/download/v0.1_aatc_testset/aatc_testset_v1_release.tar.gz}}.

We employ the objective metrics described in Section~\ref{sec:benchmarking} to evaluate the performance of the submitted systems. For the subjective evaluation, we focus on the \textbf{Acoustic Degradation} subset (150 utterances) to ensure the validity of the MOS scores. We recruited 29 listeners. Each utterance received at least 6 valid ratings after a rigorous filtering process involving ``probe samples'' (check mechanisms) to ensure data reliability.

\begin{table*}[htbp]
  \centering
  \setlength{\tabcolsep}{1.2mm}
  \caption{Quantitative comparison of all teams, the baseline, and reference signals (Clean/Noisy) on the final test set. Metrics include Word Accuracy (\textbf{WAcc}) for speech intelligibility, \textbf{DNSMOS} (SIG, BAK, OVRL) for perceptual quality, and \textbf{PESQ}. ``Params'' denotes the model parameter count in millions. The best results among the participating teams are highlighted in \textbf{bold}, and the second-best results are \underline{underlined}.}
  \label{tab:objective_metrics_results}
  \begin{tabular}{ccclccccccc}
      \toprule
      \textbf{Rank} & \textbf{Team ID} & \textbf{Type} & \textbf{Backbone} & \textbf{WAcc}$\uparrow$ & \textbf{SIG}$\uparrow$ & \textbf{BAK}$\uparrow$ & \textbf{OVRL}$\uparrow$ & \textbf{PESQ}$\uparrow$ & \textbf{Params (M)} & \textbf{Objective Score} \\
      \midrule
      - & Clean & - & - & - & 3.541 & 4.126 & 3.296 & - & - & - \\
      - & Noisy & - & - & 0.772 & 2.891 & 3.019 & 2.469 & 1.906 & - & - \\
      \midrule
      1 & T099 & D & MP-SENet~\cite{lu2023mpsenet} & \textbf{0.811} & 3.300 & \textbf{4.092} & 3.054 & \textbf{2.557} & 6.63 & \textbf{60.00} \\
      2 & T082 & Hybrid & SEMamba~\cite{chao2024semamba}, NCSNPPv2~\cite{richter2025sbvesde} & 0.779 & \textbf{3.466} & 4.069 & \textbf{3.191} & 2.165 & 7.80 & \underline{59.00} \\
      3 & T002 & D & TFGridNet~\cite{wang2023tf} & 0.779 & \underline{3.398} & 4.070 & \underline{3.135} & 2.317 & 9.42 & 58.00 \\
      4 & T081 & D & DyTSwiG-Mamba~\cite{DyTSwiG-Mamba2024} & 0.766 & 3.362 & 4.072 & 3.098 & 2.420 & 1.55 & 57.00 \\
      5 & T146 & D & MP-SENet~\cite{lu2023mpsenet} & \underline{0.788} & 3.207 & 3.977 & 2.932 & \underline{2.529} & 6.63 & 56.00 \\
      6 & T056 & D & PrimeK-Net~\cite{lin2025primek} & 0.762 & 3.388 & \underline{4.087} & 3.127 & 2.345 & 1.41 & 55.00 \\
      7 & T066 & Hybrid & PGUSE~\cite{zhang2025pguse} & 0.777 & 3.321 & 4.011 & 3.049 & 2.483 & 5.37 & 54.00 \\
      8 & T112 & D & CrossNet~\cite{kalkhorani2024crossnet} & 0.785 & 3.318 & 3.947 & 3.012 & 2.157 & 3.54 & 53.00 \\
      9 & T068 & D & MP-SENet~\cite{lu2023mpsenet} & 0.764 & 3.347 & 4.000 & 3.081 & 2.348 & 4.10 & 52.00 \\
      10 & T130 & D & Mamba-SEUNet~\cite{wang2025mambaseunet} & 0.759 & 3.304 & 4.079 & 3.051 & 2.258 & 6.32 & 51.00 \\
      11 & T032 & D & MP-SENet~\cite{lu2023mpsenet} & 0.722 & 3.302 & 3.934 & 3.001 & 2.161 & 2.21 & 50.00 \\
      12 & T115 & D & MP-SENet~\cite{lu2023mpsenet} & 0.720 & 3.324 & 3.897 & 2.989 & 2.013 & 6.29 & 48.00 \\
      13 & T076 & D & DeepFilterNet3~\cite{schroeter2023deepfilternet3} & 0.711 & 3.139 & 3.957 & 2.868 & 2.065 & 2.10 & 47.00 \\
      14 & T071 & D & DCCRN~\cite{hu2020dccrn} & 0.635 & 3.119 & 3.740 & 2.784 & 1.850 & 3.51 & 46.00 \\
      15 & T079 & D & BSRNN~\cite{yu2023bsrnn} & 0.731 & 3.441 & 3.905 & 3.111 & 2.205 & 26.35 & 41.00 \\
      16 & T052 & Hybrid & AnyEnhance~\cite{zhang2025anyenhance}, TFGridNet~\cite{wang2023tf} & 0.692 & 3.353 & 4.012 & 3.072 & 2.134 & 86.29 & 35.00 \\
      17 & T108 & G & AnyEnhance~\cite{zhang2025anyenhance} & 0.635 & 3.340 & 4.070 & 3.063 & 1.772 & 82.18 & 34.00 \\
      18 & T132 & G & AnyEnhance~\cite{zhang2025anyenhance} & 0.634 & 3.357 & 4.062 & 3.088 & 1.738 & 82.68 & 33.00 \\
      19 & T117 & G & AnyEnhance~\cite{zhang2025anyenhance} & 0.624 & 3.365 & 4.075 & 3.099 & 1.741 & 82.18 & 32.00 \\
      20 & T104 & G & AnyEnhance~\cite{zhang2025anyenhance} & 0.620 & 3.404 & 4.069 & 3.134 & 1.711 & 82.18 & 30.00 \\
      21 & T009 & D & Mossformer2~\cite{zhao2024mossformer2} & 0.676 & 3.191 & 3.917 & 2.883 & 1.981 & 55.26 & 29.00 \\
      22 & T017 & D & Dasheng Denoiser~\cite{sun2025dashengdenoiser} & 0.727 & 3.177 & 3.865 & 2.886 & 1.777 & 138.50 & 29.00 \\
      23 & T093 & G & AnyEnhance~\cite{zhang2025anyenhance} & 0.631 & 3.224 & 4.065 & 2.966 & 1.640 & 80.34 & 28.00 \\
      \rowcolor{gray!20}
      24 & Baseline & G & AnyEnhance~\cite{zhang2025anyenhance} & 0.618 & 3.290 & 4.050 & 3.031 & 1.672 & 82.18 & 27.00 \\
      24 & T004 & D & HifiGAN~\cite{kong2020hifigan} & 0.441 & 3.308 & 3.958 & 2.998 & 1.260 & 50.98 & 26.00 \\
      25 & T111 & G & AnyEnhance~\cite{zhang2025anyenhance} & 0.002 & 2.449 & 3.689 & 1.895 & 1.232 & 82.18 & 26.00 \\
      \bottomrule
  \end{tabular}
\end{table*}

\subsection{Overview of Submitted Systems}
\label{subsec:systems}

\begin{figure}[!htbp]
  \centering
  \includegraphics[width=0.95\linewidth]{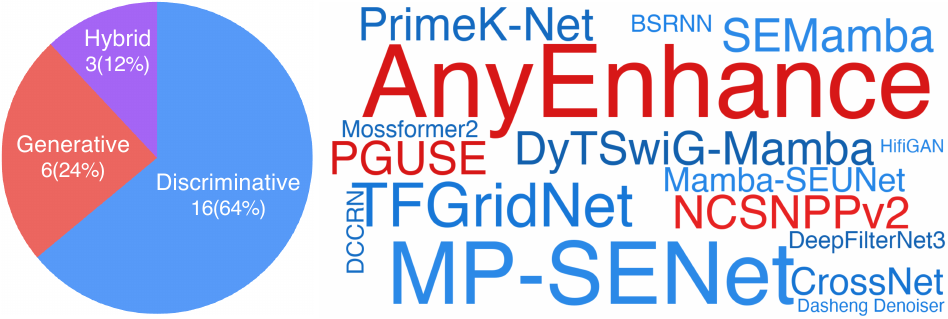}
  \caption{Overview of the model architectures adopted by the top 25 teams. \textbf{Left:} The quantitative distribution of model paradigms, showing that discriminative models are the most common choice. \textbf{Right:} A word cloud visualization of the specific backbones and components, highlighting the high frequency of ``AnyEnhance'' and ``MP-SENet''. Colors in the table correspond to the word cloud: \textcolor[HTML]{4E79A7}{Blue} for Discriminative and \textcolor[HTML]{E15759}{Red} for Generative components.}
  \label{fig:model_distribution_cloud}
\end{figure}

We received \textbf{25} valid submissions among \textbf{104} registered teams. As illustrated in Fig.~\ref{fig:model_distribution_cloud} (Left), the solutions can be categorized into three distinct paradigms: Discriminative (64\%), Generative (24\%), and Hybrid (12\%). Fig.~\ref{fig:model_distribution_cloud} (Right) further provides a word cloud visualization of the specific backbones employed, revealing the community's preference for established T-F domain architectures and the emerging interest in state-space models.

\begin{itemize}
    \item \textbf{Discriminative Models (D):} This category represents the predominant choice, adopted by 16 teams.
    \textbf{Time-Frequency (T-F) domain} architectures remain the gold standard, with \textbf{MP-SENet}~\cite{lu2023mpsenet} and \textbf{TF-GridNet}~\cite{wang2023tf} being the most popular backbones.
    Notably, there is a rising trend towards efficiency-oriented designs: several teams (e.g., T081, T130) incorporated \textbf{Mamba/State Space Models}~\cite{gu2024mamba} (e.g., DyTSwiG-Mamba, Mamba-SEUNet) to reduce computational complexity while maintaining receptive fields. As detailed in Table~\ref{tab:objective_metrics_results}, these models are characterized by their lightweight footprint, with most discriminative entries containing fewer than 10M parameters.

    \item \textbf{Generative Models (G):} Six teams opted for generative approaches. The majority built upon the official baseline, \textbf{AnyEnhance}~\cite{zhang2025anyenhance}, a masked generative model.
    Unlike the compact discriminative models, these systems are significantly larger, typically exceeding 80M parameters (e.g., Baseline at 82.18M). This increased size is primarily due to the heavy dependencies on neural audio codecs (DAC) and large transformer backbones required for token prediction.

    \item \textbf{Hybrid Models (H):} Three teams explored hybrid architectures that combine discriminative feature extraction with generative refinement. A representative example is T082, which integrates \textbf{SEMamba}~\cite{chao2024semamba} (a selective state-space model) with \textbf{NCSNPPv2}~\cite{richter2025sbvesde} (a score-based diffusion component). This design aims to leverage the stability of discriminative learning and the perceptual quality of generative priors. However, this hybrid nature often results in variable parameter counts.
\end{itemize}

In summary, the competition submissions highlight a distinct dichotomy in model scaling: discriminative approaches strive for extreme lightness ($<$10M) suitable for edge deployment, while generative approaches prioritize capacity ($>$80M) to model complex distributions.

\subsection{Overall Performance and Efficiency Analysis}
\label{sec:overall_performance}

Table~\ref{tab:objective_metrics_results} presents the quantitative comparison of the top-performing teams, while Fig.~\ref{fig:rank_vs_params} visualizes the relationship between model size and final ranking. A joint analysis reveals two critical insights defining the current state of speech restoration:

\begin{figure}[!htbp]
  \centering
  \includegraphics[width=0.8\linewidth]{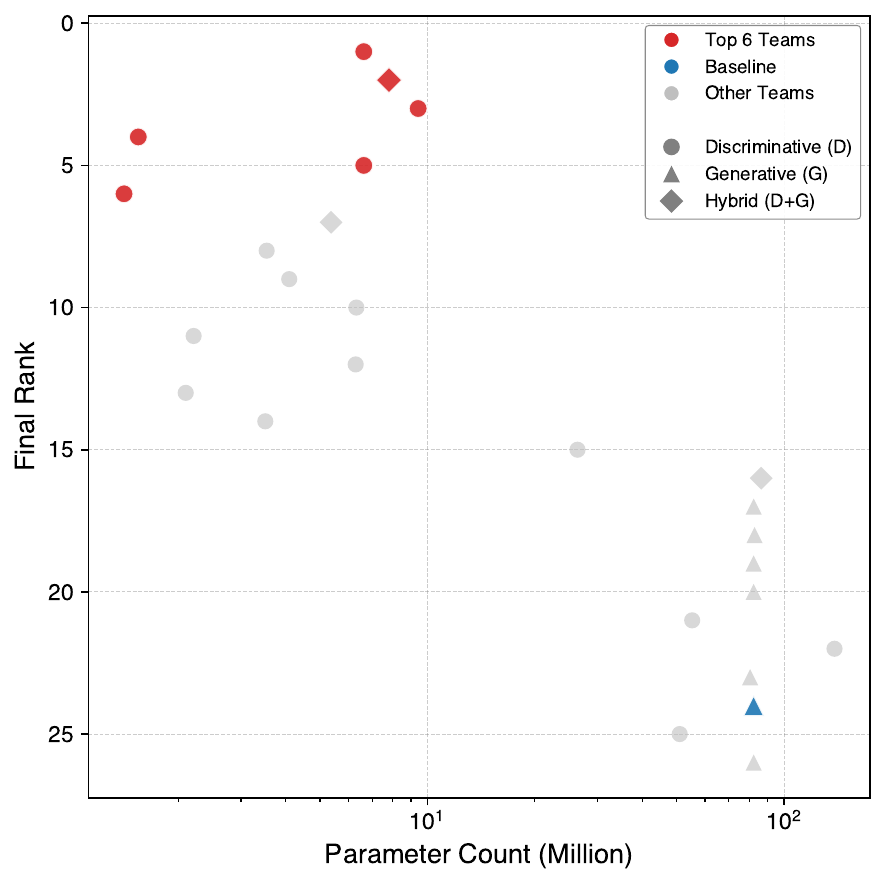}
  \caption{Comparison of model parameter efficiency and final ranking. The x-axis represents the parameter count (in millions) on a logarithmic scale, and the y-axis represents the final competition rank. Colors distinguish the team categories, while marker shapes indicate the model architecture types (Discriminative, Generative, or Hybrid).}
  \label{fig:rank_vs_params}
\end{figure}

\subsubsection{The Efficiency Frontier:}
Contrary to the prevailing trend in deep learning where "larger is better," our results highlight a distinct \textbf{low-resource Pareto Front}. As shown in Fig.~\ref{fig:rank_vs_params}, the top 6 participating teams all cluster in the top-left corner, utilizing fewer than 10M parameters.
\begin{itemize}
    \item \textbf{Lightweight Dominance:} The winning team, T099 (MP-SENet), achieved state-of-the-art performance with only 6.63M parameters. Even more strikingly, T056 (PrimeK-Net) ranked 6th with an extremely compact footprint of 1.41M parameters, outperforming models 50$\times$ its size.
    \item \textbf{The Generative Burden:} In contrast, generative models (marked as triangles in Fig.~\ref{fig:rank_vs_params}) mostly populate the bottom-right region. For instance, the Baseline and related entries (e.g., T108, T117) exceed 80M parameters due to the heavy computational cost of neural codec decoders and autoregressive backbones. However, this increased capacity did not translate into superior objective scores, largely because they struggled to recover fine-grained phase information required for high PESQ, and their generative hallucinations penalized WAcc.
\end{itemize}

\subsubsection{Discriminative vs. Hybrid: The Fidelity-Quality Trade-off}
A granular analysis of Table~\ref{tab:objective_metrics_results} reveals a fascinating divergence between the top-ranked discriminative model (T099) and the top-ranked hybrid model (T082).
\begin{itemize}
    \item \textbf{Fidelity (T099):} T099 dominates in metrics reflecting signal fidelity and intelligibility, achieving the highest \textbf{WAcc (0.811)} and \textbf{PESQ (2.557)}. This suggests that discriminative mapping in the T-F domain remains the most effective strategy for preserving phonemic content and waveform alignment.
    \item \textbf{Perceptual Quality (T082):} T082, which integrates a generative component, achieved the highest \textbf{SIG (3.466)} and \textbf{OVRL (3.191)} scores. This indicates that incorporating generative priors helps synthesize ``perceptually'' pleasing speech.
\end{itemize}
However, T082's lower PESQ (2.165 vs. 2.557) and WAcc (0.779 vs. 0.811) compared to T099 imply a trade-off: hybrid models gain perceptual smoothness at the cost of signal fidelity. This duality sets the stage for the breakdown analysis in the following sections.

\begin{figure}[!ht]
  \centering
  \includegraphics[width=0.7\linewidth]{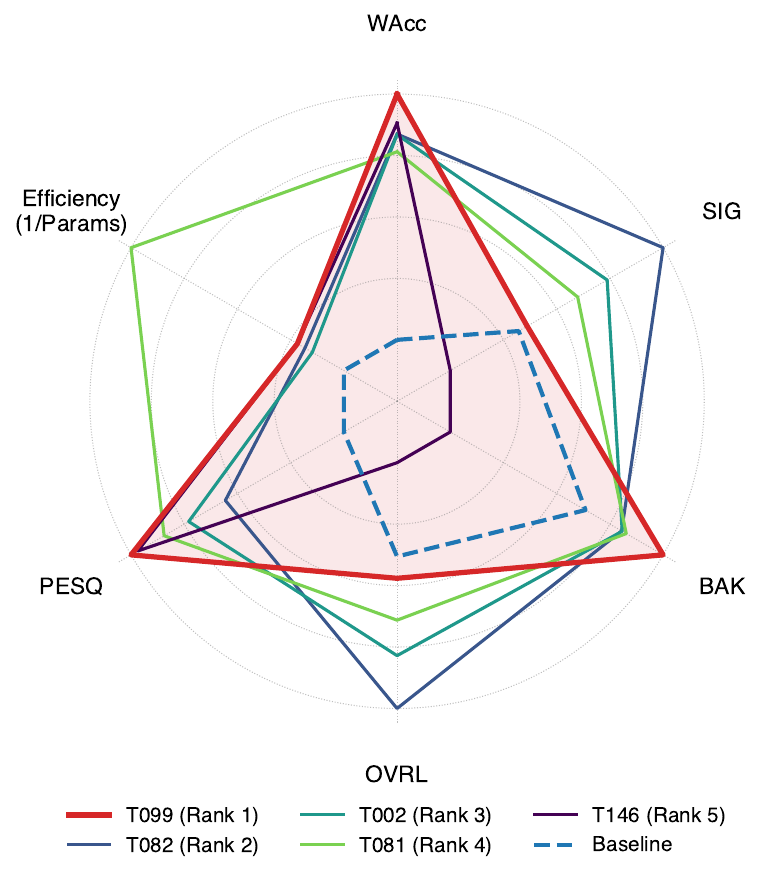}
  \caption{Radar chart comparing the Top 5 teams and the baseline across six dimensions. All metrics are normalized to the [0, 1] range for visualization. Note that for \textit{Efficiency}, a larger value indicates fewer parameters (calculated as $1/\text{Params}$), while for other metrics, higher values indicate better performance.}
  \label{fig:radar_chart_comparison}
\end{figure}

To visualize these trade-offs, Fig.~\ref{fig:radar_chart_comparison} presents the normalized metrics across six dimensions. The rank 1 system T099 (Red) exhibits the largest enclosed area with a balanced shape, demonstrating holistic robustness across intelligibility, fidelity, and efficiency. In contrast, the hybrid model T082 (Blue) shows a distinct ``skewed'' profile: it extends furthest on the SIG axis (perceptual quality) but retreats significantly on PESQ and Efficiency. This visual confirmation reinforces that, under current constraints, lightweight discriminative models offer a superior compromise compared to the specialized but computationally heavier hybrid approaches.

\begin{table*}[htbp]
  \centering
  \caption{Detailed performance comparison across three distortion types: Acoustic Degradation, Codec Distortion, and Secondary Processing Artifacts. The best results among participating teams are highlighted in \textbf{bold}.}
  \label{tab:type_analysis}
  \begin{tabular}{cllccccc}
      \toprule
      \textbf{Distortion Type} & \textbf{Rank} & \textbf{Team} & \textbf{WAcc}$\uparrow$ & \textbf{SIG}$\uparrow$ & \textbf{BAK}$\uparrow$ & \textbf{OVRL}$\uparrow$ & \textbf{PESQ}$\uparrow$ \\
      \midrule
      \multirow{9}{*}{Acoustic Degradation}
      & - & Clean & - & 3.526 & 4.126 & 3.281 & - \\
      & - & Noisy & 0.731 & 2.408 & 2.026 & 1.837 & 1.306 \\
      \cmidrule(lr){2-8}
      & - & Baseline & 0.530 & 3.221 & 4.047 & 2.976 & 1.505 \\
      \cmidrule(lr){2-8}
      & 1 & T099 & \textbf{0.790} & 3.218 & \textbf{4.088} & 2.978 & \textbf{2.337} \\
      & 2 & T082 & 0.756 & \textbf{3.491} & 4.034 & \textbf{3.191} & 1.960 \\
      & 3 & T002 & 0.743 & 3.375 & 4.069 & 3.112 & 2.108 \\
      & 4 & T081 & 0.730 & 3.313 & 4.070 & 3.051 & 2.165 \\
      & 5 & T146 & 0.754 & 3.116 & 3.925 & 2.839 & 2.331 \\
      \midrule
      \multirow{9}{*}{Codec Distortion} 
      & - & Clean & - & 3.556 & 4.110 & 3.300 & - \\
      & - & Noisy & 0.989 & 3.547 & 4.091 & 3.282 & 4.112 \\
      \cmidrule(lr){2-8}
      & - & Baseline & 0.929 & 3.537 & 4.099 & 3.274 & 2.638 \\
      \cmidrule(lr){2-8}
      & 1 & T099 & 0.990 & 3.546 & 4.109 & 3.289 & \textbf{4.324} \\
      & 2 & T082 & 0.977 & 3.528 & 4.127 & 3.285 & 3.619 \\
      & 3 & T002 & 0.986 & 3.560 & 4.109 & 3.304 & 4.117 \\
      & 4 & T081 & \textbf{0.994} & \textbf{3.567} & \textbf{4.129} & \textbf{3.321} & 4.159 \\
      & 5 & T146 & 0.993 & 3.534 & 4.091 & 3.272 & 4.186 \\
      \midrule
      \multirow{9}{*}{Secondary Processing Artifacts}
      & - & Clean & - & 3.554 & 4.136 & 3.315 & - \\
      & - & Noisy & 0.743 & 3.254 & 3.961 & 2.978 & 1.758 \\
      \cmidrule(lr){2-8}
       & - & Baseline & 0.607 & 3.312 & 4.042 & 3.041 & 1.489 \\
      \cmidrule(lr){2-8}
      & 1 & T099 & \textbf{0.764} & 3.330 & \textbf{4.100} & 3.084 & 2.110 \\
      & 2 & T082 & 0.735 & \textbf{3.368} & 4.088 & \textbf{3.113} & 1.808 \\
      & 3 & T002 & 0.738 & 3.320 & 4.047 & 3.057 & 1.788 \\
      & 4 & T081 & 0.728 & 3.353 & 4.046 & 3.078 & 2.013 \\
      & 5 & T146 & 0.750 & 3.298 & 4.032 & 3.024 & \textbf{2.115} \\
      \bottomrule
  \end{tabular}
\end{table*}

\subsection{Breakdown Analysis by Distortion Type}
\label{sec:distortion_analysis}

To analyze the specific capabilities and limitations of different modeling paradigms, we conduct a granular performance analysis across the three distinct distortion categories as detailed in Table~\ref{tab:type_analysis}. This breakdown reveals a fundamental trade-off between signal fidelity and perceptual hallucination.

\subsubsection{Acoustic Degradation: The Fidelity-Smoothness Trade-off}
This category, representing standard denoising and dereberveration tasks, highlights the divergence in optimization goals. 
Discriminative models, exemplified by T099, prioritize signal fidelity, achieving the highest WAcc (0.790) and PESQ (2.337). These models effectively act as precise non-linear filters, preserving phonemic edges essential for intelligibility. In contrast, hybrid approaches like T082 excel in spectral reconstruction, achieving the highest SIG score (3.491) and OVRL (3.191). However, this comes at a cost: T082's PESQ (1.960) is notably lower than T099's, suggesting that while the generated spectra are smooth and noise-free (pleasing to DNSMOS), the phase coherence and temporal alignment with the reference are compromised during the generative refinement process.

\subsubsection{Codec Distortion: The Curse of Reconstruction Bias}
The results on Codec Distortion provide the strongest evidence for the ``Reconstruction Bias'' inherent in generative models. It is crucial to note that the input signals, despite mp3 compression, maintain high intelligibility and phase alignment (Input WAcc 0.989, PESQ 4.112). 
\begin{itemize}
    \item Discriminative Success: Top-performing discriminative models (T099, T081, T146) correctly identify the high SNR nature of the input. They approximate an \textit{identity mapping} while subtly correcting quantization artifacts, thereby preserving or slightly enhancing the high input PESQ (e.g., T099 improves PESQ to 4.324).
    \item Generative Failure: Conversely, generative systems struggle to be conservative. The Baseline model drastically degrades the input, dropping WAcc to 0.929 and PESQ to 2.638. Even the top-tier hybrid model T082 degrades the input PESQ from 4.112 to 3.619. This indicates that without explicit constraints, generative models tend to aggressively resynthesize the speech based on learned priors rather than repairing the existing, high-quality structure, leading to unnecessary phase mismatches and artifact introduction.
\end{itemize}

\subsubsection{Secondary Processing Artifacts: The Generalization Bottleneck}
This category represents the most challenging distortion type, characterized by upstream artifacts that are highly non-predictable and non-linear. All models exhibit a performance drop compared to other categories. However, discriminative models (T099) again demonstrate superior robustness in retaining content (WAcc 0.764) compared to hybrid counterparts. The hybrid model T082, while achieving the highest perceptual scores (SIG 3.368), suffers from the lowest PESQ (1.808) among the top 5. This suggests that when faced with unfamiliar artifacts, generative components are prone to hallucination, masking artifacts by synthesizing plausible but phase-incoherent speech textures, which satisfy reference-free metrics but deviate from the ground truth.

\begin{figure*}[!htbp]
  \centering

  \subfloat[SIG\label{fig:corr_sig}]{
    \includegraphics[width=0.19\linewidth]{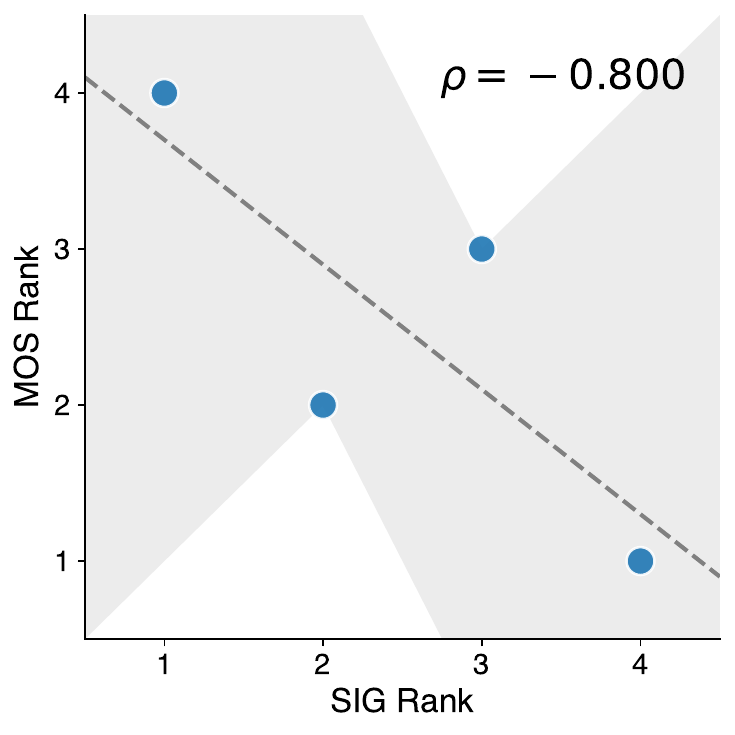}
  }
  \subfloat[BAK\label{fig:corr_bak}]{
      \includegraphics[width=0.19\linewidth]{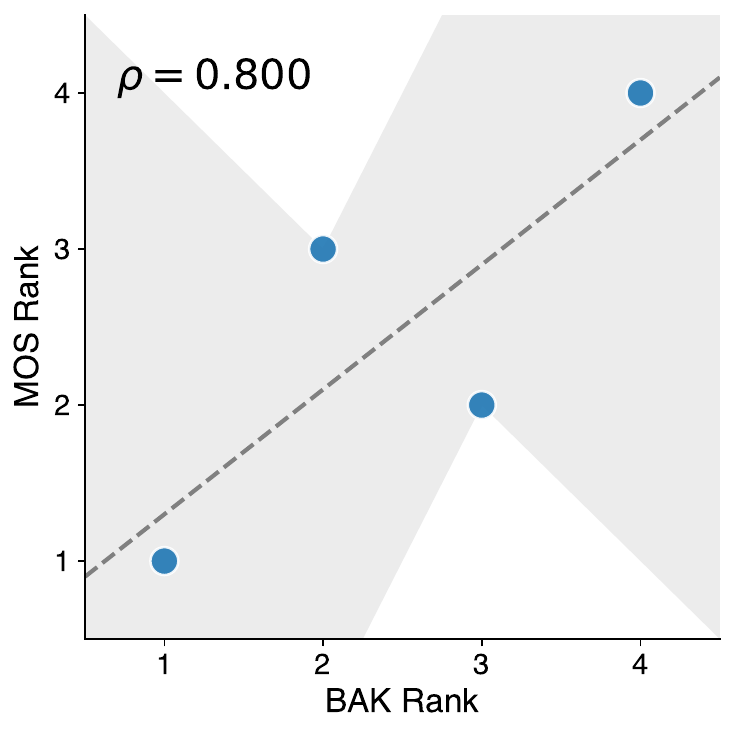}
  }
  \subfloat[OVRL\label{fig:corr_ovrl}]{
      \includegraphics[width=0.19\linewidth]{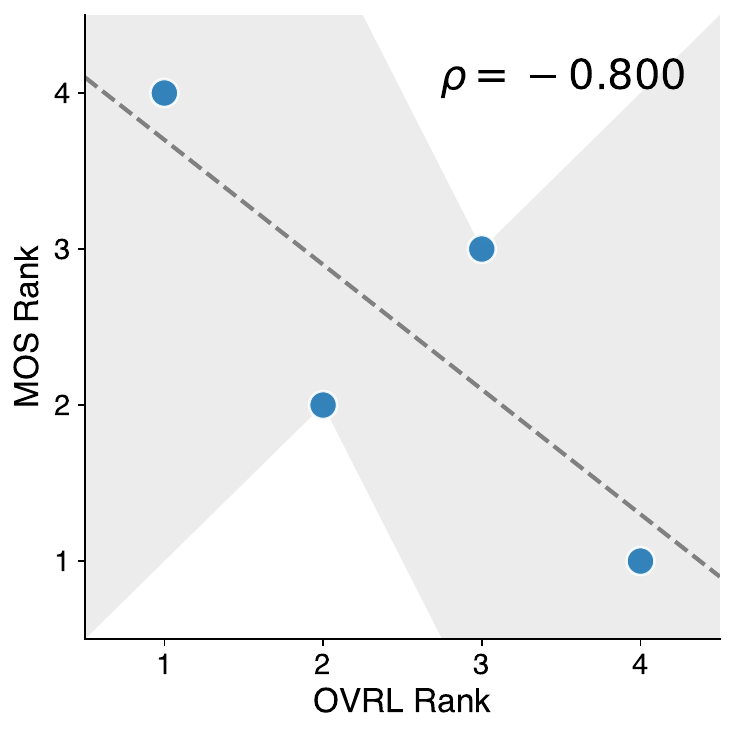}
  }
  \subfloat[PESQ\label{fig:corr_pesq}]{
      \includegraphics[width=0.19\linewidth]{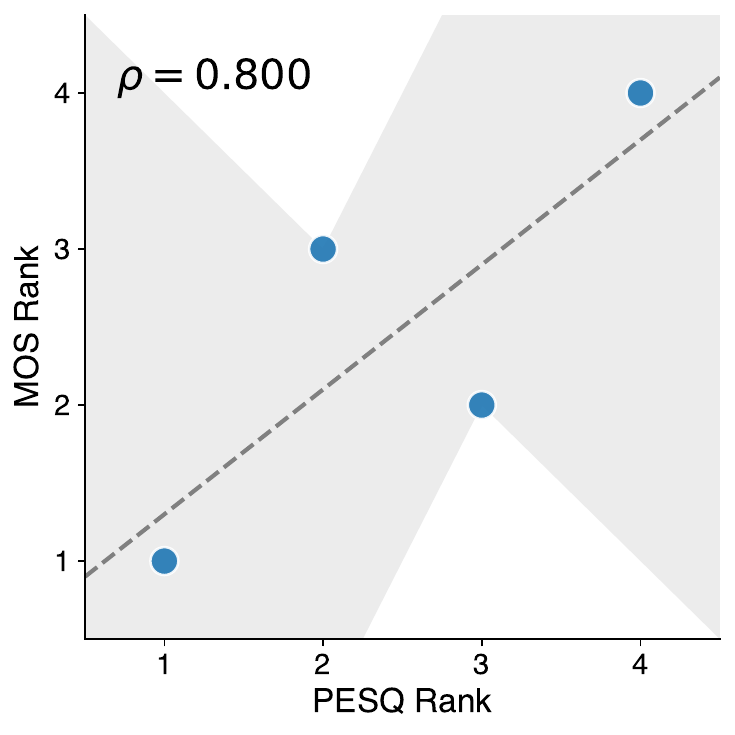}
  }
  \subfloat[WAcc\label{fig:corr_wacc}]{
      \includegraphics[width=0.19\linewidth]{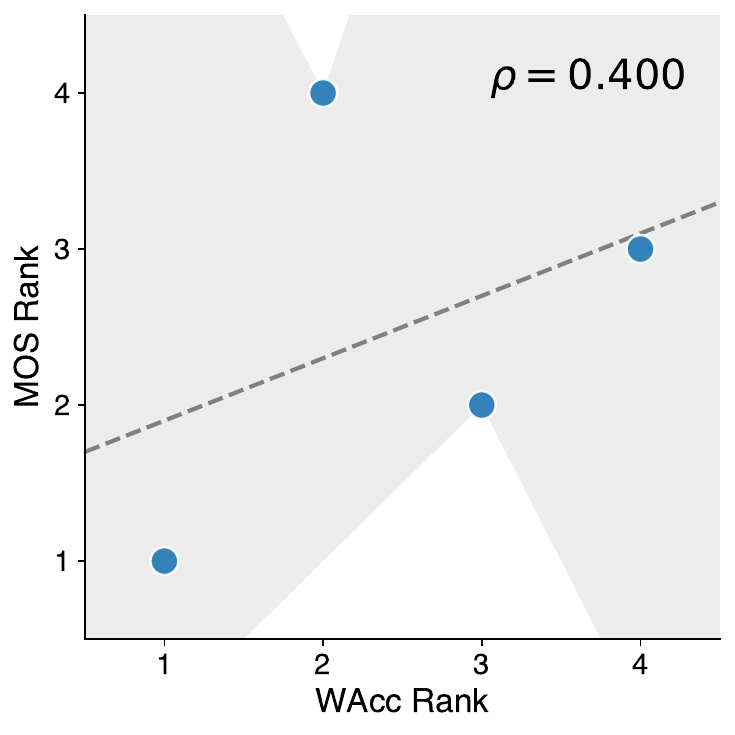}
  }
 
  \caption{Rank-rank correlation analysis between subjective MOS and five objective metrics. Each point represents a participating team. The dashed line indicates the linear regression fit, and the shaded area represents the 95\% confidence interval. $\rho$ denotes the Spearman's rank correlation coefficient. A higher $\rho$ indicates better alignment with human perception.}
  \label{fig:correlation_analysis}
\end{figure*}

\begin{figure}[!ht]
  \centering
  \includegraphics[width=0.8\linewidth]{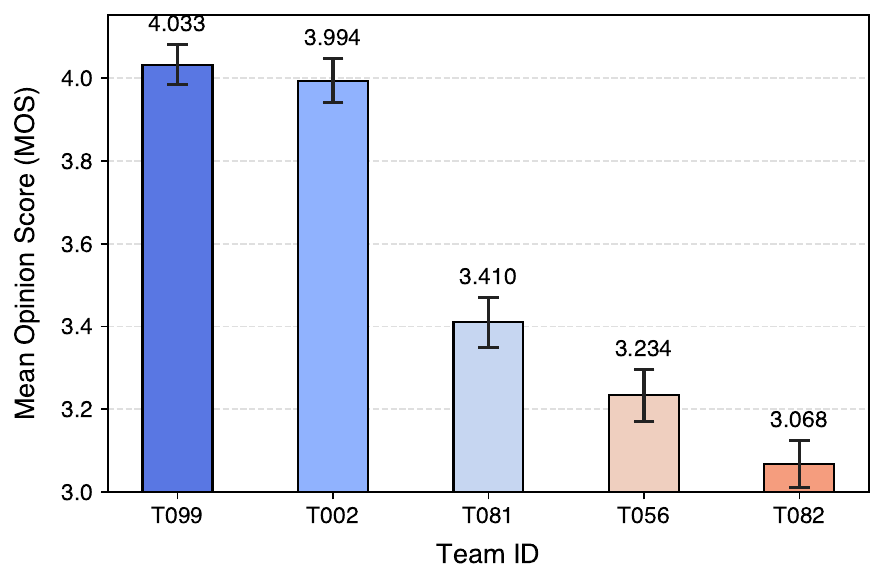}
  \caption{MOS scores of the top 5 teams in the final round, with 95\% confidence interval.}
  \label{fig:mos_scores}
\end{figure}

\subsection{Divergence between Objective and Subjective Quality}
\label{subsec:metric_divergence}

A critical discrepancy emerges when scrutinizing the alignment between objective metrics and human perception. While discriminative models (e.g., T099) maintain consistency across metrics, hybrid/generative approaches expose a severe ``Metric Gap''.

\subsubsection{The Uncanny Valley of Spectral Smoothness}
As visualized in Fig.~\ref{fig:mos_scores}, the discriminative model T099 dominates the subjective evaluation with a MOS of \textbf{4.033}. In stark contrast, the hybrid model T082, despite incorporating advanced generative priors, ranks lowest among the top five systems (MOS \textbf{3.068}).

This result presents a paradox when cross-referenced with Table~\ref{tab:type_analysis}:
\begin{itemize}
    \item \textbf{The DNSMOS Illusion:} T082 achieved the highest DNSMOS SIG score (\textbf{3.491}) in the competition, significantly outperforming T099 (\textbf{3.218}). Current reference-free metrics appear to over-reward the perfectly smooth spectral envelopes synthesized by generative models, interpreting the absence of noise as ``high quality.''
    \item \textbf{Human Reality:} Listeners significantly penalized T082. Qualitative feedback indicates that while the background was silent, the speech exhibited an ``artificial robotic timbre'' and subtle phase discontinuities—artifacts distinct from additive noise that DNSMOS fails to penalize. This confirms that T082 falls into an \textit{uncanny valley}: statistically clean, yet perceptually unnatural.
\end{itemize}

\subsubsection{Rank-Rank Correlation Analysis}
To quantify this misalignment, we calculated the Spearman's rank correlation coefficient ($\rho$) between objective metrics and human MOS on the acoustic subset (Fig.~\ref{fig:correlation_analysis}). The results overturn conventional wisdom in three aspects:

\begin{enumerate}
    \item \textbf{Inversion of Reference-Free Metrics (SIG/OVRL):}
    Most alarmingly, DNSMOS SIG and OVRL exhibit a strong \textbf{negative correlation ($\rho=-0.8$)} with human ratings. This implies that for hybrid systems, optimizing for higher DNSMOS scores currently drives the model \textit{away} from human preference. While the metric correctly assesses noise suppression (BAK $\rho=0.8$), it is fundamentally unable to distinguish between natural speech texture and over-smoothed, synthesized artifacts.
    
    \item \textbf{The Resurgence of Fidelity Metrics (PESQ):}
    Contrary to the trend of disregarding intrusive metrics for generative tasks, PESQ demonstrates a strong \textbf{positive correlation ($\rho = 0.800$)}. This suggests that in the context of universal restoration, human listeners prioritize \textbf{signal fidelity} (waveform consistency) over aggressive hallucination. T099's high PESQ (2.337) successfully predicted its subjective superiority, whereas T082's lower PESQ (1.960) correctly signaled its perceptual degradation.
\end{enumerate}

The current generation of reference-free metrics has not yet adapted to the artifacts introduced by generative and hybrid models. They act as a ``smoothness detector'' rather than a ``naturalness detector,'' necessitating the development of next-generation metrics sensitive to generative hallucinations.

\begin{figure*}[!htbp]
  \centering

  \subfloat[Acoustic Degradation Case 1: T082 exhibits severe spectral smearing (red box) yet achieves the highest SIG score (3.57), illustrating metric misalignment. The Baseline suffers from phoneme loss.\label{fig:casestudy_degraded_73}]{
    \includegraphics[width=0.9\linewidth]{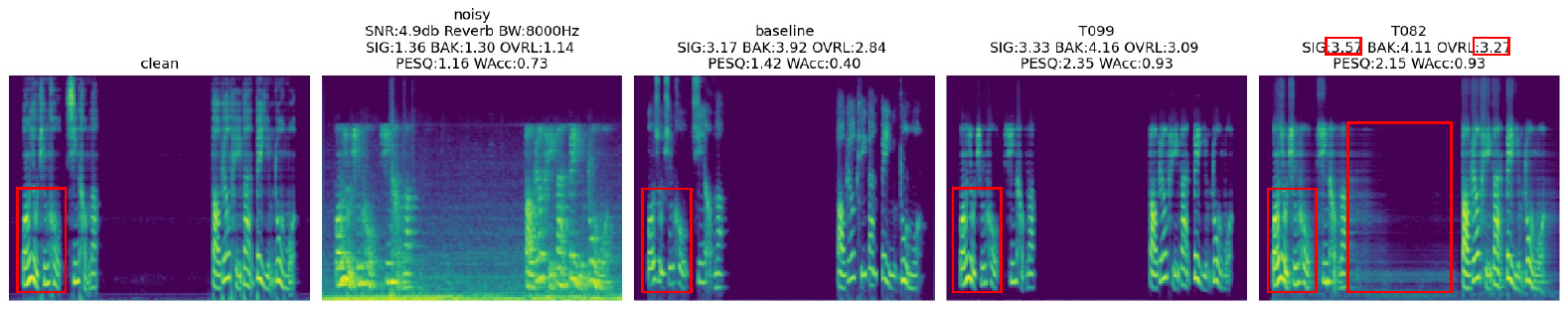}
  }
  \\
  \subfloat[Acoustic Degradation Case 2: Comparison of harmonic preservation. T099 maintains sharp formant structures, whereas T082 introduces temporal blurring artifacts in high frequencies.\label{fig:casestudy_degraded_263}]{
      \includegraphics[width=0.9\linewidth]{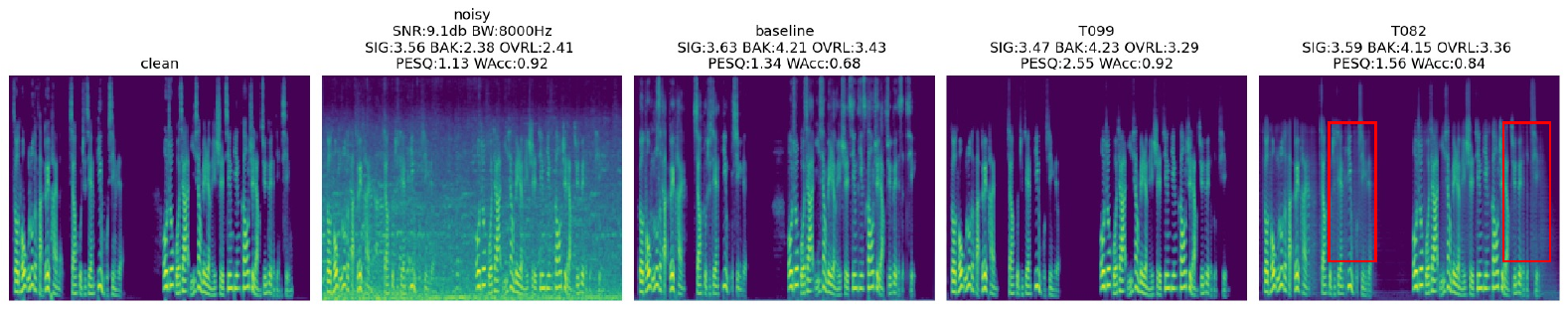}
  }
  \\
  \subfloat[Secondary Processing Artifacts Case 1 (Input: TF-GridNet): All models effectively sharpen the blurred input. Note that T082 still introduces vertical smearing artifacts.\label{fig:casestudy_generated_782}]{
      \includegraphics[width=0.9\linewidth]{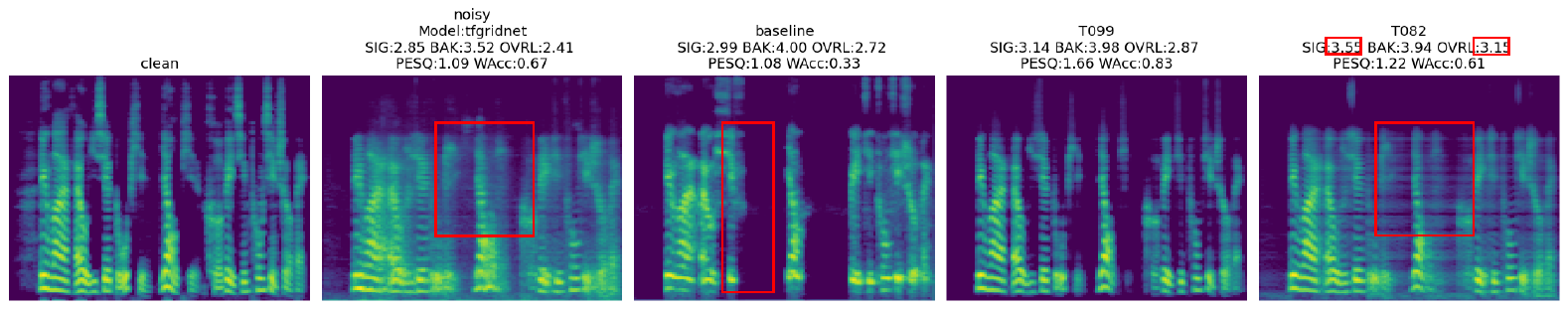}
  }
  \\
  \subfloat[Secondary Processing Artifacts Case 2 (Input: LLaSE-G1): The upstream model introduces a "spectral block" hallucination. Discriminative (T099) and Hybrid (T082) models preserve this artifact, while the Generative Baseline successfully removes it based on speech priors.\label{fig:casestudy_generated_959}]{
      \includegraphics[width=0.9\linewidth]{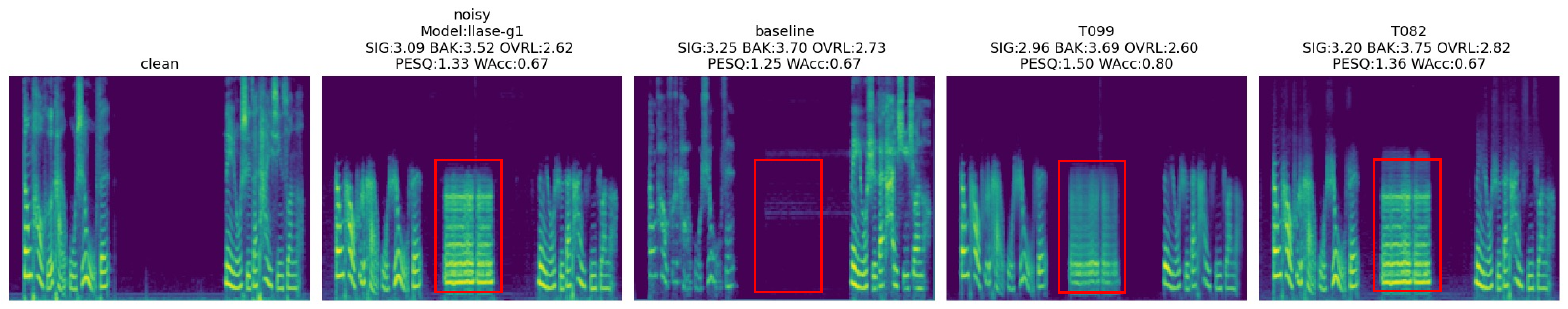}
  }
 
  \caption{Spectrogram visualization of the top-performing Discriminative (T099), Hybrid (T082), and Generative (Baseline) models across different distortion types. Red boxes highlight critical artifacts: \textbf{smearing} in Hybrid models (a, c) and \textbf{hallucinations} from upstream models (d). Metrics (SIG/BAK/OVRL/PESQ/WAcc) are annotated above each spectrogram. }
  \label{fig:casestudy}
\end{figure*}

\subsection{Case Study and Visualization}
\label{subsec:casestudy}

To visually validate our quantitative findings, we inspect spectrograms from the three representative paradigms: Discriminative (T099), Generative (Baseline), and Hybrid (T082). Fig.~\ref{fig:casestudy} presents samples from Acoustic Degradation and Secondary Processing Artifacts.

\subsubsection{Acoustic Degradation} Fig.~\ref{fig:casestudy_degraded_73} and Fig.~\ref{fig:casestudy_degraded_263} highlight the trade-off between restoration fidelity and spectral smoothness. 
\begin{itemize}
    \item \textbf{Generative Instability:} The Baseline model, while removing noise, often damages the harmonic structure, leading to broken formants (red box in Fig.~\ref{fig:casestudy_degraded_73}) and lower intelligibility (WAcc 0.40).
    \item \textbf{Metric Hacking in Hybrid Models:} A critical observation is the behavior of the hybrid model T082. As marked by the red boxes in Fig.~\ref{fig:casestudy_degraded_73} and \ref{fig:casestudy_degraded_263}, T082 exhibits severe \textbf{spectral smearing} and temporal blurring, likely artifacts from the diffusion refinement process. Paradoxically, T082 achieves the highest DNSMOS SIG scores (e.g., 3.57 vs. T099's 3.33 in Case 1). This visual evidence confirms that current reference-free metrics over-penalize the residual noise in sharp discriminative outputs (T099) while failing to detect the unnatural ``oversmoothing'' artifacts introduced by generative components.
\end{itemize}

\subsubsection{Secondary Processing Artifacts} In this track, models must restore speech processed by upstream enhancement models.
\begin{itemize}
    \item \textbf{Restoration of Discriminative Artifacts:} In Fig.~\ref{fig:casestudy_generated_782}, the input is blurred by TF-GridNet. All three models successfully sharpen the spectrum. However, T082 again introduces vertical smearing artifacts (red box), further confirming its tendency towards temporal inconsistency.
    \item \textbf{The Potential of Generative Priors:} Fig.~\ref{fig:casestudy_generated_959} presents a unique scenario where the upstream model (LLaSE-G1) introduced a \textbf{spectral hallucination}—a synthetic block of energy in the silence region (see Input spectrogram). Interestingly, both the discriminative (T099) and hybrid (T082) models treated this artifact as valid speech content and preserved it. In contrast, the generative baseline \textit{completely removed} this hallucination. This suggests that pure generative models, driven by strong speech priors, have the unique potential to identify and reject semantically improbable structures that discriminative models—bound by input fidelity—fail to filter out.
\end{itemize}


\section{Conclusion}
This paper presented a comprehensive retrospective of the CCF AATC 2025 Speech Restoration Challenge, establishing a benchmark for universal blind speech restoration. Through a rigorous evaluation of 25 participating systems, we demonstrated that lightweight discriminative models currently offer the superior trade-off between fidelity and efficiency, outperforming large-scale generative approaches that suffer from reconstruction bias in high-SNR scenarios. Most critically, our analysis exposed a severe misalignment between objective metrics and human perception, highlighting that widely used reference-free metrics may actively penalize naturalness in the era of generative enhancement. We hope these insights serve as a catalyst for the development of adaptive restoration architectures and perception-aligned evaluation protocols.

\bibliographystyle{IEEEtran}
\bibliography{refs}

\end{document}